\documentclass[%
 reprint,
nofootinbib,
 amsmath,amssymb,
aps,
floatfix,
]{revtex4-2}
\usepackage{graphicx}
\usepackage{dcolumn}
\usepackage{bm}
\usepackage{xcolor}
\usepackage{comment}
\usepackage[utf8]{inputenc}

\newcommand*\diff{\mathop{}\!\mathrm{d}}
\newcommand{\virgolette}[1]{``#1''}
\newcommand{\redradius}{z}    
\newcommand\mathperiod{\,.}
\newcommand\mathcomma{\,,}
\count\footins = 1000
\begin{document}

\preprint{APS/123-QED}

\title{Impact of a modified Entropy-Area law on Schwarzschild-de Sitter metric}

\author{L. Marchetti}
\email[]{luca.marchetti@phd.unipi.it}
\affiliation{Arnold Sommerfeld Center for Theoretical Physics, Ludwig-Maximilians-Universit\"at München Theresienstrasse 37, 80333 M\"unchen, Germany, EU}
\affiliation{Università di Pisa, Lungarno Antonio Pacinotti 43, 56126 Pisa, Italy, EU}
\affiliation{Istituto Nazionale di Fisica Nucleare sez. Pisa, Largo Bruno Pontecorvo 3, 56127 Pisa, Italy, EU}

\author{G. Cella}
\email[]{giancarlo.cella@pi.infn.it}
\affiliation{Istituto Nazionale di Fisica Nucleare sez. Pisa Largo B. Pontecorvo 3, 56127 Pisa (Italy)}

\date{\today}

\begin{abstract}
Based on the perspective that continuum gravitational physics is an emergent quantum gravitational phenomenon, and that spacetime thermodynamic is the natural langauge in which it can be described, we derive a modified Schwarzschild-de Sitter geometry in static coordinates from quantum gravity inspired logarithmic corrections to the entropy-area law. The resulting quantum corrections to classical geometry can be encoded in explicitly radius dependent black hole and cosmological constant parameters. By considering the pure black hole case we study linear perturbations on this modified background, obtaining corrections to the black hole quasi-normal modes suppressed by the square of the ratio between the Planck length and the Schwarzschild radius.
\end{abstract}

\maketitle

\section{Introduction}

One of the most intriguing feautures of gravity is that the dynamics of spacetime can be given a thermodynamic interpretation \cite{Jacobson:1995ab}. This discovery, followed by many other works aiming to better understand the deep connection between gravity and thermodynamics both in specific cases \cite{Cai:2005ra,Cai:2006rs,Padmanabhan:2002sha,Zhang:2013tca} and on more general ground \cite{,Hayward:1998ee,Hayward:1997jp,Hayward:1993mw,Jacobson:2015hqa,Padmanabhan:2013nxa,Padmanabhan:2007en,Kothawala:2010bf,Chirco:2009dc,Eling:2008af,PhysRevLett.96.121301,Kothawala:2010bf} (see for instance \cite{Padmanabhan:2009vy} for a review), naturally led to the see gravity as an emergent phenomenon. 

A similar perspective has been developed after decades of theoretical efforts devoted to solve the Quantum Gravity (QG) problem from different approaches. The continuum description of spacetime is, from this point of view, indeed emergent, arising from the collective behavior of some fundamental \virgolette{atoms of space}. These degrees of freedom are not simply quantized continuum fields; they do not have any spatiotemporal meaning nor they have a direct geometric interpretation \cite{Oriti:2018dsg}. Some examples are the spin networks of Loop Quantum Gravity (LQG) \cite{Ashtekar:2004eh, Perez:2012wv,rovelli_2004,Rovelli1998} (though they were introduced first within a canonical quantization of the gravitational field), the piecewise-flat simplicial geometries of lattice QG approaches \cite{Hamber:2009mt, Ambjorn:2012jv}, the quanta of Group Field Theories (GFTs) \cite{Oriti:2011jm, Krajewski:2012aw, Oriti:2017ave}, causal sets~\cite{Dowker:aza}, and possibly the underlying degrees of freedom of String Theory \cite{Blau:1900zza}.   

If gravity is in fact an emergent phenomenon, probing its thermodynamic properties could help us understanding its \virgolette{microscopic structure} \cite{Padmanabhan:2010xe}. Moreover, if QG theories are able to predict quantum corrections to the classical thermodynamic variables, then thermodynamic equations, which we expect to hold both in classical and quantum regime, may be used to describe corrections to the usual gravitational dynamics, i.e., to the Einstein equations. This is the basic principle that we subscribe to in this paper.

A key role in the thermodynamic approach to gravity is played by horizon entropy. Interestingly enough, a lot of effort was put into trying to compute corrections to the \virgolette{classical} Bekenstein-Hawking area law \cite{bekenstein,Bardeen1973,Hawking1975} due to quantum geometry \cite{Rovelli:1996dv,Ashtekar:1997yu,Kaul:2000kf,Ghosh:2004rq,Domagala:2004jt,Meissner:2004ju,Chatterjee:2003su,Engle:2011vf,Livine:2005mw} and from quantum and thermal fluctuations \cite{Solodukhin:1997yy,Mann:1997hm,Kastrup:1997iu,Das:2001ic,Gour:2003jj,Mukherji:2002de,Chatterjee:2003uv}, obtaining, in general, a functional form given by
\begin{equation}\label{eqn:entropy1}
    \mathcal{S}=\frac{\mathcal{A}}{4L_{\text{pl}}^2}+b\log\frac{\mathcal{A}}{4L_{\text{pl}}^2}+\mathcal{O}\left( \frac{\mathcal{A}}{4L_{\text{pl}}^2}\right)^{-1},
\end{equation}
where $\mathcal{A}$ is the horizon area. Unfortunately, while there is broad agreement on the logarithmic form of these next-to-the-leading-order corrections, the factor b seems to be very model dependent \cite{Carlip:2014pma}. However, it is interesting to notice that conformal field theory arguments \cite{Carlip:2000nv}, as well as computations in LQG \cite{Meissner:2004ju,Engle:2011vf,Livine:2005mw} and GFT \cite{Oriti:2018qty}, all seem to suggest a value $b=-3/2$.
These quantum corrections could be used, in the spirit described above, to obtain corrections to the standard gravitational dynamics. 

A first difficulty of this program is that typically the corrections are obtained in the spherically symmetric case, and there is no a priori reason to assume them to be true for more general spacetimes. Hence one has in principle no idea of how he entropy functional should be corrected.

Another issue is that even if one believes Equation~(\ref{eqn:entropy1}) to capture the supposed entropy modifications, the standard procedure developed by Jacobson~\cite{Jacobson:1995ab} makes use of local Rindler horizons. Being these local, one has not a clear notion of what the area of the horizon should be. 

Given these difficulties, one is led to consider first the spherically symmetric case, in order to understand first what happens in this simplified picture and to hopefully find some (perhaps) observable corrections. Indeed, a first tentative in this direction has been performed in \cite{Cai:2008ys}, where an effective cosmological dynamics has been obtained strikly resembling the effective dynamics obtained in the Loop Quantum Cosmology setting \cite{Taveras:2008ke} (see \cite{Bojowald:2008zzb} for a review). Another possibility, explored in \cite{Alonso-Serrano:2020dcz,Alonso-Serrano:2020pcz} involves the implementation of such logarithmic corrections at the level of local causal diamonds, obtaining modified Einstein equations.

In this paper, we will only consider the very simple situation of a spherically symmetric static case, and we derive from a thermodynamic approach the corrections to the Schwarzschild--de Sitter solution induced by equation~\eqref{eqn:entropy1} by imposing the validity of fundamental thermodynamic equations for each spherically symmetric surface of a spatial hypersurface.
In section~\ref{sec:SDS} we will apply specifically this program of obtaining possible quantum corrections to the gravitational dynamics in a very simple case, the Schwarzschild-de Sitter case, finding a modified solution. In section~\ref{sec:PERT} we will study the linear perturbation around this modified background in the  Schwarzschild case. Finally we draw some conclusions in section~\ref{sec:CONCL}.

\section{Modified geometry from entropy corrections}
\label{sec:SDS}

In order to obtain a modified dynamics from quantum corrections to the entropy functional, it is necessary to recast the gravitational dynamics in terms of thermodynamic quantities. We will do this by following the approach in \cite{Padmanabhan2013}, which we will briefly review here.

As a starting point, one considers a spacetime which is assumed to be foliated by space-like leaves $\Sigma_t$ characterized by a constant value of a scalar field $t(x^a)$. Coordinates $y^\alpha$ can be defined on the leaves by defining a congruence of time--like curves parameterized by $t$, each curve connecting events with the same value of $y^\alpha$.

The unit normal to the $\Sigma$ is given by $u_{a}=-N\nabla_{a}t$ and we can write the induced metric on $\Sigma$ as
\begin{equation}
h_{ab}\equiv g_{ab}+u_a u_b\mathperiod
\end{equation}
Note that $h^a_b$ is the projector on the leaves. A coordinate change can be written as
\begin{equation}
    dx^a = \left( N u^a +N^a \right) dt + x^a_{,\alpha} dy^\alpha\mathcomma
\end{equation}
where $N_a$ is tangent to $\Sigma_t$, namely $h^a_b N_a = N_b$. The \virgolette{time-developement vector} $\zeta^a = N u^a + N^a$ tells how the coordinates change with the time if $y^\alpha$ is keep fixed.

The relevant vector here is $\xi^a\equiv N u^a$, the component of $\zeta^a$ which is normal to $\Sigma_t$. As $g_{0\alpha}=N_\alpha$, when $g_{0\alpha}=0$ then $\zeta^a=\xi^a$. If the spacetime is static $\xi^a$ can be identified with a time-like Killing vector. 

\subsection{Noether charge of $\xi^a$ and surface heat content}

Because of diffeomorphism invariance, there exists a Noether charge associated to any vector field in spacetime, generating a corresponding diffeomorphism. If we call $J^a(\xi)$ the conserved current associated to the above vector $\xi^a$, one can show that
\begin{equation}\label{eqn:noethercurrent}
u_{a}J^a(\xi)=\frac{1}{8\pi L_{\text{pl}}^2}D_{\alpha}(Na^\alpha)\,,
\end{equation}
where $a^i\equiv u^j\nabla_j u^i$ is the acceleration and $D_{i}a^i=D_{\alpha}a^\alpha=\nabla_ia^i-a^2$. where $D_{i}$ is the covariant derivative on the constant $t$ slices. By definition, $Na_{i}\equiv Nu^j\nabla_j u_i=\xi^j\nabla_j u_i$ measures the change in velocity along the vector $\xi^i$. One can also show that for $a_{i}$ the following identity holds:
\begin{equation}
N a_{i}=h^j_i\nabla_j N\,.
\end{equation}
Now, consider a $t=\text{const.}$ region $\mathcal{V}$, and let us integrate equation \eqref{eqn:noethercurrent} multiplied by the correct volume element $\sqrt{h}\diff^3x$ on it: we find
\begin{equation}
\int_{\mathcal{V}}\sqrt{h}\diff^3x u_aJ^a(\xi)=\int_{\partial V}\frac{\sqrt{\sigma}\diff^2x}{8\pi L_{\text{pl}}^2}(Nr_\alpha a^\alpha)\,.
\end{equation}
where $r_\alpha$ it the unitary normal to the boundary surface $\partial V$.

Now, consider, in particular, a surface characterized by $N(t,\mathbf{x})=\text{const.}$ inside the $t=\text{const.}$ hypersurface. Then $r_\alpha$ is the normal to this particular surface, so that we can write $r_\alpha\propto D_{\alpha}N$, or equivalently as $r_{i}\propto h^j_i\nabla_jN$, meaning that $r_i$ and $Na_{i}$ are in the same direction. We can thus normalize this vector by choosing $r_{\alpha}=\epsilon a_\alpha/a$, with $\epsilon=\pm 1$, so that $r_{\alpha}$ is always pointing outward. As a consequence, we see that $Nr_{\alpha}a^\alpha=\epsilon Na=\epsilon(h^{ij}\nabla_i N\nabla_jN)^{1/2}$. Now, following \cite{Padmanabhan2013}, we identify the quantities appearing at the right-hand-side of the above equation as follows:
\begin{itemize}
\item First, we notice that the quantity $Na/(2\pi)$ can be interpreted as the (Tolman redshifted) local Davies-Unruh temperature of observes with $u_{a}=-N\delta^0_a$, i.e.\ of observers whose motion is normal to the hypersurface $t=\text{const.}$ and whose acceleration is $a$ with respect to freely falling observers. 
\item Second, we interpret $s=\sqrt{\sigma}/(4L_{\text{pl}}^2)$ as an entropy density. This follows because any small patch of the boundary we are considering can be considered as a correspondingly small cross-section of the local Rindler horizon for a local Rindler observer accelerating with acceleration $a$ (who thus associate a temperature $T_{\text{loc}}\equiv Na/(2\pi)$ to this surface).
\end{itemize}
In terms of these quantities, we can then write
\begin{equation}\label{eqn:integratedcurrent}
\int_{\mathcal{V}}\sqrt{h}\diff^3x\,u_{\alpha}J^\alpha(\xi)=\epsilon\int_{\partial \mathcal{V}}\diff^2 x\, Ts\,,
\end{equation}
so that, as long as the boundary is defined as a surface characterized by $N(t,\mathbf{x})=\text{const.}$ in a constant-$t$ hypersurface, the Noether charge for the vector $\xi$, integrated on a physical region of space, is equal to the boundary enthalpy.

\subsection{Gravitational dynamics and thermodynamics}

Up to this point, however, the gravitational field equations have not been used. When this is done, one can use equation \eqref{eqn:integratedcurrent} to recast the integrated gravitational dynamics as \cite{Padmanabhan2013}
\begin{align}
&\frac{1}{16\pi L_{\text{pl}}^2}\int_{\mathcal{V}}\sqrt{h}\diff^3x\, u_{a}g^{ij}\mathcal{L}_{\xi}N^a_{ij}\nonumber\\
&\quad=\epsilon\int_{\partial \mathcal{V}}\diff^2x\, Ts-\int_{\mathcal{R}}\diff^3x N\sqrt{h}u^au^b(\tilde{T}_{ab})\,,
\end{align}
where $\tilde{T}_{ab}=T_{ab}-Tg_{ab}/2$, with $T\equiv T^a_a$, and $\mathcal{L}_{\xi}N^a_{ij}$ is the Lie derivative along $\xi$ of the quantity
\begin{equation*}
N^a_{bc}=-\Gamma^a_{bc}+\frac{1}{2}(\Gamma^d_{bd}\delta^a_c+\Gamma^d_{cd}\delta^a_{b})\,,
\end{equation*}
which can be interpreted as the canonical momentum of the quantity $f^{ab}=\sqrt{-g}g^{ab}$. In particular, one can show that 
\begin{equation*}
\sqrt{h}u_{a}g^{ij}\mathcal{L}_{\xi}N^a_{ij}=-h_{ab}\mathcal{L}_{\xi} p^{ab}\,,
\end{equation*}
where $p_{ab}$ is the canonically conjugate momentum to the projection tensor $h_{ab}$. In the following, we will be interested to a very particular case of the above equation: when the spacetime is static, $\xi^a$ is a Killing vector, and the left-hand-side of the above equation vanishes, giving
\begin{equation}
\label{eqn:equilibrium}
\epsilon\int_{\partial \mathcal{V}}\diff^2x\, Ts=\int_{\mathcal{V}}\diff^3x N\sqrt{h}u^au^b(\tilde{T}_{ab})\mathperiod
\end{equation}
In~\cite{Padmanabhan2013}, this equation is interpreted as \emph{holographic equipartition}, imposing that the number of degrees of freedom on the bulk and on the boundary is the same. Here, however, an interpretation in terms of thermodynamic variables will be more convenient. 
\subsection{Modified Schwarzschild-de Sitter geometry}

Equation~\eqref{eqn:equilibrium}, having an immediate thermodynamic interpretation, can be used as a starting point for the programme of obtaining modified geometries from a modified entropy functional, at least in the simplest cases. 

In the following we will focus only on a cosmological constant-like energy momentum tensor, thus characterized by a constant energy density: $T^a_b=\rho\delta^a_b$. As a consequence, $\tilde{T}^a_b=-\rho\delta^a_b$. 

Let us consider the construction performed as before, and let us adopt $t$ as time variable; further, let us adapt our coordinates to the spherical symmetry shown by the system. We thus make the ansatz
\begin{equation}
\diff s^2=-f(r)\diff t^2+h(r)\diff r^2+r^2\diff\Omega^2\,.
\end{equation}
Thus, the boundary we are interested to, is a $r=\text{const}.$ in a $t=\text{const.}$ hypersurface, i.e., a $2$-sphere of radius $r$. 

Given the symmetry properties of the system we can write the equilibrium equation \eqref{eqn:equilibrium} (choosing $\epsilon=-1$) as
\begin{equation}
T(r)S(r)=-\frac{4}{3}\pi r^3\rho \sqrt{f(r)h(r)} + \Upsilon\,,
\end{equation}  
where $S(r)=\int_{\partial S_2(r)}\diff^2 x\, s(r)$. Here we used $u^au^b\tilde{T}_{ab}=-\rho u^au_a=\rho$, and we added a constant $\Upsilon$ related to a singular density at the center. In this coordinate system, furthermore,
\begin{align}
\label{eq:temperature}
T &=\frac{Na}{2\pi} \nonumber \\
  &=\frac{1}{2\pi} \sqrt{h^{ij}\nabla_iN\nabla_jN} \nonumber \\
  & =\frac{1}{4\pi}(fh)^{-1/2}f'(r)\,.
\end{align} 
It is useful to reparametrize $h(r)=f^{-1}(r) e^{g(r)}$, so that 
\begin{align}
	\label{eqn:conservationwithg}
TS &=-\frac{4}{3}\pi r^3\rho e^{g/2}+\Upsilon\mathcomma \\
T &=\frac{f'(r)e^{-g/2}}{4\pi}\,.
\end{align}

In order to determine both the functions $f$ and $g$, and hence determine the full form of the metric in this coordinate system, we need one more relation among these quantities, besides the equation above. This can be obtained as done in~\cite{Zhang:2013tca}, by imposing that the dynamical surface gravity $\kappa=\frac{1}{2}\nabla_a\nabla^a r$ matches the \virgolette{acceleration surface gravity} $\kappa(r)=Na=2\pi T$. By definition, we have
\begin{align*}
\kappa&=\frac{1}{2}\left(\partial_a\nabla^a r+\Gamma^a_{a\mu}\nabla^\mu r\right) \nonumber \\
&=\frac{1}{2}\left(\partial_a (g^{a\mu}\nabla_\mu r)+\Gamma^a_{a\mu}g^{\mu\nu}\nabla^\nu r\right) \nonumber \\
&=\frac{1}{2}\left(\partial_a (g^{a\mu}\partial_\mu r)+\Gamma^a_{a\mu}g^{\mu\nu}\partial_\nu r\right) \nonumber \\
&=\frac{1}{2}\left(\partial_{r}g^{rr}+\Gamma^a_{ar}g^{rr}\right)=\frac{1}{2}\left(\partial_{r}h^{-1}+\Gamma^a_{ar}h^{-1}\right)\mathperiod
\end{align*}

We now evaluate the Christoffel symbols 
\begin{align*}
\Gamma^0_{0r}&=f^{-1}\partial_r f\mathcomma\\
\Gamma^r_{rr}&=h^{-1}\partial_r h\mathcomma
\end{align*}
finally obtaining
\begin{align}
\label{eq:surfgrav1}
\kappa&=\frac{h^{-1}}{4}\left(f^{-1}\partial_r f-h^{-1}\partial_r h\right)\mathperiod
\end{align}
In particular, we notice immediately that this definition agrees with $\kappa=\partial_r f/2$ when $f=h^{-1}$. 

Equating the two resulting definitions of surface gravity, from equation~\eqref{eq:surfgrav1} and equation~\eqref{eq:temperature}, we find
\begin{equation*}
\frac{1}{2}(fh)^{-1/2}f'(r)=\frac{h^{-1}}{4}\left(f^{-1}\partial_r f-h^{-1}\partial_r h\right),
\end{equation*}
which, once rewritten in terms of $f$ and $g$, becomes
\begin{equation}\label{eqn:equalitysurfacegravities}
f'\left(1-e^{g/2}\right)=\frac{1}{2}fg'\,.
\end{equation}
The general solution can be written explicitly in the form
\begin{equation}
	g = \log \left( \frac{f}{f+C}   \right)^2
\end{equation}
where $C$ is an integration constant. We note that by choosing $C=0$ we obtain $g=0$, which correspond to $hf=1$, the usual condition for a spherically symmetric solution in the Schwarzschild form. This condition in fact guarantees the correct asymptotic behaviour when a cosmological constant is not present, and we will stick to it.

We could now insert in~\eqref{eqn:conservationwithg} the modified form of the entropy functional~\eqref{eqn:entropy1}, 
\begin{equation}
\label{eq:modifiedS}
S\simeq \frac{\pi r^2}{L_{\text{pl}}^2}\left(1+\frac{b}{\pi}\frac{L_{\text{pl}}^2}{r^2}\log\frac{\pi r^2}{L_{\text{pl}}^2}\right)\,,
\end{equation}
and solve the resulting equation. This is quite easy to do, and is described in Appendix~\ref{app:calcolointegrato}.

However, in the spirit of interpreting our fundamental equation as a thermodynamic one, we prefer to consider the derivative of equation~\eqref{eqn:conservationwithg} with respect to $r$ as our fundamental dynamical equation, describing the change of enthalpy across different spherical surfaces due to the presence of a matter source\footnote{This is similar to the approach in~\cite{Zhang:2013tca}.}. 

Recalling that, from the above arguments, $g=0$, for $\pi r^2\gg L_{\text{pl}}^2$, the radial derivative of equation \eqref{eqn:conservationwithg} becomes
\begin{equation}
f''_{\rho,b}(r)\left(r^2+\frac{b}{\pi}\log\pi r^2\right)+f'_{\rho,b}(r)\left(2r+\frac{b}{2 r}\right)=-16\pi \rho r^2\,,
\end{equation}
where for the sake of notation, we have put $L_{\text{pl}}=1$ (the right powers of $L_{\text{pl}}$ will be reinstated below). Let us neglect the second term going as $r^{-1}$, which is very small in the limit of large $r$, and let us look first for a solution of the homogeneous equation. This can be easily integrated one time to get $f'_b(r)=c_1K_b(r)$, with
\begin{equation}\label{eqn:derivativefb}
K_b(r)=\exp\left(-\int^r_1\diff z\frac{2z}{z^2+\frac{b}{\pi}\log\pi z^2}\right)\,,
\end{equation}
and we notice that, for large $r$, $K_b(r)$ goes as $r^{-2}$ plus other subleading terms. Next, we look for a particular solution. Consider an ansatz of the form $f'_{\rho,b}(r)=A_{\rho,b}(r)K_b(r)$. The quantity $A_{\rho,b}(r)$ must then satisfy the equation
\begin{equation*}
A'_{\rho,b}(r)=-16\pi \rho\frac{r^2 }{r^2+(b/\pi)\log \pi r^2} K_b^{-1}(r)\,.
\end{equation*}
Its formal integration then gives $A(\rho,b)(r)$, and we can write the most general solution for $f'$ as
\begin{equation*}
f'_{\rho,b}(r)=\left(A_{\rho,b}(r)+c_1\right)K_b(r)\,.
\end{equation*}
We can now write analythically the solution for $f(r)$ by treating perturbatively the logarithmic terms in the integral for $K_b(r)$ (since for each $y\ge 1$ we have $\log (\pi y^2)/(\pi y^2)\ll 1$). In this way we just need to solve the following integral:
\begin{align*}
&\int_1^z\diff y\frac{2}{y}\left(1-b\frac{\log(\pi y^2)}{\pi y^2}\right)\\
&\quad=2\log z\left(1+\frac{b}{\pi z^2}\right)-\frac{b}{\pi}(1+\log \pi)\left(1-\frac{1}{z^{2}}\right)\nonumber\\
&\quad\simeq 2\log z\left(1+\frac{b}{\pi z^2}\right)-\frac{b}{\pi}(1+\log \pi)\,,
\end{align*}
where the last term has again been neglected because, in our approximations, it is much smaller than all the other ones. So, we obtain
\begin{equation*}
K_b(r)\simeq \frac{1}{r^2}-\frac{2b}{\pi}\frac{\log r}{r^4}\,.
\end{equation*}
Now, if we perform the same expansion also for $A'_{\rho,b}$, we obtain
\begin{equation*}
A'_{\rho,b}\simeq -16\pi \rho \left(r^2+\frac{b}{\pi^2} \log\pi\right)\,,
\end{equation*}
from which we would conclude that
\begin{equation*}
A_{\rho,b}=-\frac{16\pi \rho}{3}r^3\left(1+\frac{b\log\pi}{3\pi^2r^2}\right)+c_2\,,
\end{equation*}
where the integration constant can be reabsorbed into the constant $c_1$. Now, in the usual spirit of neglecting corrections of order $\mathcal{O}(r^{-2})$, we can safely neglect the second term in round brackets and we find
\begin{equation*}
f'_{\rho,b}\simeq \left(-\frac{16\pi \rho}{3}r^3+c_1\right)r^{-2}\left(1-\frac{2b}{\pi}\frac{\log r}{r^2}\right)\,.
\end{equation*}
This equation is easily integrated and, neglecting again term of order $r^{-2}$, we find
\begin{equation*}
f_{\rho,b}(r)\simeq c_2-\frac{c_1}{r}\left(1-\frac{2b}{3\pi}\frac{\log r}{r^2}\right)-\frac{8\pi \rho}{3}r^2\left(1-\frac{2b}{\pi}\frac{\log^2r}{r^2}\right)\,.
\end{equation*}
We finally fix the integration constants so that 
\begin{equation}
f_{\rho,b=0}=1-\frac{r_{S}}{r}-\frac{8\pi\rho}{3} r^2\,,
\end{equation} 
which means $c_{2}=1$ and $c_{1}=r_S$. By reinstating powers of $L_{\text{pl}}$, we finally conclude
\begin{widetext}
\begin{align}
\label{eqn:fbr}
f_{\rho,b}(r)&\simeq 1-\frac{r_S}{r}\left(1-\frac{2b}{3\pi}\frac{L_{\text{pl}}^2}{r^2}\log\frac{r}{L_{\text{pl}}}\right)
-\frac{8\pi \rho L_{\text{pl}}^2}{3} r^2\left(1-\frac{2b}{\pi}\frac{L_{\text{pl}}^2}{r^2}\log^2\frac{r}{L_{\text{pl}}}\right).
\end{align}
\end{widetext}
The same result is obtained in appendix~\ref{app:calcolointegrato} with an alternative procedure. We see that quantum corrections to the metric function are suppressed by the very small ratio $L_{\text{pl}}^2/r^2$ for $r$ larger than the Schwarzschild radius $r_S$ (but within the range of validity of the coordinate system chosen, determined by $f_{\rho,b}(r)>0$). Moreover, we notice that the form of $f_{\rho,b}(r)$ naturally suggests to define effective $r$-dependent black hole and cosmological constant parameters as
\begin{subequations}
\begin{align}
    r_S(r)\equiv r_S\left(1-\frac{2b}{3\pi}\frac{L_{\text{pl}}^2}{r^2}\log\frac{r}{L_{\text{pl}}}\right)\mathcomma\\
    \rho(r)\equiv \rho \left(1-\frac{2b}{\pi}\frac{L_{\text{pl}}^2}{r^2}\log^2\frac{r}{L_{\text{pl}}}\right)\mathperiod
\end{align}
\end{subequations}
Notice also that the position of horizons is modified by the quantum corrections. For instance, in the pure de Sitter case, the new horizon is given by
\begin{equation}
    \tilde{r}^2_\rho=\frac{3}{8\pi \rho L_{\text{pl}}^2}\left(1+\frac{b}{\pi}\frac{8\pi\rho L_{\text{pl}}^4}{3}\log^2\frac{3}{8\pi\rho L_{\text{pl}}^4}\right)\mathperiod
\end{equation}
Therefore, the associated Hubble parameter $H^2=\tilde{r}^{-2}_\rho$ can be perturbatively written as
\begin{align}
    H^2&=\frac{8\pi\rho L_{\text{pl}}^2}{3}\left(1-\frac{b}{\pi}\frac{8\pi\rho L_{\text{pl}}^4}{3}\log^2\frac{3}{8\pi\rho L_{\text{pl}}^4}\right)\nonumber\\
    &=\frac{8\pi\rho L_{\text{pl}}^2}{3}\left(1-\frac{\rho}{\rho_c}\log^2\frac{3}{8\pi\rho L_{\text{pl}}^4}\right)\mathcomma
\end{align}
where
\begin{equation}
    \rho_c\equiv \frac{3}{8b L_{\text{pl}}^4}\mathperiod
\end{equation}
This behavior (modulo the logarithmic factor) is the same obtained in \cite{Cai:2008ys}, though the definition of $\rho_c$ differs from \cite{Cai:2008ys} by a factor of $2$. 

The same horizon analysis can of course be performed in the pure Scwarzschild case. This will be discussed in detail in the next section.
\section{Linear perturbations}
\label{sec:PERT}
Among the many interesting black hole physics phenomena that can be studied, a particularly important one is how a black hole reacts to small external perturbances. When perturbed, the response of black holes is characterized by a set of quasi-normal modes, which have been thoroughly investigated both at the theoretical (see~\cite{Berti2009} for a review) and experimental~\cite{BuonannoQNM,LIGOQNM} level. In this section, we study black hole perturbations on the modified background of equation in~\eqref{eqn:fbr} (with $\rho=0$) by means of a WKB approximation in two simple cases: first, in Section~\ref{sec:scalarperturbations} we study the behavior of a scalar perturbation, which already shows the most important features of the physics involved; then, in Section \ref{sec:axialpert}, we consider axial tensor perturbations, whose fundamental evolution equations share with the scalar ones remarkable mathematical similarities.

\subsection{Scalar perturbations}\label{sec:scalarperturbations}

Let us first study the dynamics of a simple scalar field on the modified background. After the spherical decomposition
\begin{equation}
    \phi\equiv \sum_{\ell m}\frac{u(r,t)}{r}Y_{lm}(\theta,\varphi)\,,
\end{equation}
where $Y_{lm}$ are spherical armonics, the equation of motion
\begin{equation*}
   0= \Box\phi=\frac{1}{\sqrt{-g}}\partial_\mu\left(\sqrt{-g}g^{\mu\nu}\partial_\nu\phi\right)
\end{equation*}
becomes
\begin{equation}\label{eqn:equationmotion1}
    f_{b}\partial_r\left(f_{b}\partial_r u_{\ell m}\right)-\partial_t^2u_{\ell m}-V_{\ell}(r)u_{\ell m}=0\,,
\end{equation}
where
\begin{align}\label{eqn:vll}
    V_{\ell}(r)&\equiv f_{b}\left[\frac{\partial_r f_{b}}{r}+\frac{\ell(\ell+1)}{r^2}\right]\nonumber\\
    &=\left(1-\frac{r_S(r)}{r}\right)\frac{r_S(r)-rr_S'(r)+r\ell(\ell+1)}{r^3}\,,
\end{align}
and where we have redefined the factor $f_{\rho,b}(r)\vert_{\rho=0}\equiv f_b(r)$ as
\begin{equation}\label{eqn:frredefined}
    f_{b}(r)\equiv (1-r_S(r)/r)\,,\\
\end{equation}
Since there are two length scales involved in the defnition of $f_{b}(r)$, it is useful to define the following parameters:
\begin{equation}
\epsilon_S\equiv L_{\text{pl}}^2/r_S^2\,,\qquad \redradius\equiv r/r_S\,.
\end{equation}
The first of course determines the perturbative expansion, since we are supposing to consider macroscopic black-holes, while the second one measures the radial coordinate in unites of the Schwartzschild radius. In term of these two quantities, we have that
\begin{align}
f_{b}(\redradius)&= 1-\frac{1}{\redradius}\left(1-\epsilon_S\frac{c_b}{\redradius^2}\log\frac{\redradius}{\sqrt{\epsilon_S}}\right)\,,\\
V_{\ell}(\redradius)&=\frac{f_{b}(\redradius)}{r_{S}^2}\left[\frac{\partial_{\redradius}f_{b}(\redradius)}{\redradius}+\frac{\ell(\ell+1)}{\redradius^2}\right],
\end{align}
where $c_{b}\equiv 2b/3\pi$ is a parameter supposed to be of order $1$, which is the standard scenario where quantum logarithmic corrections appear. Written in terms of these new variables it is easy to perturbatively find the zero of $f_{b}(\redradius)$. In fact, a simple computation shows that the new zero is approximately at 
\begin{equation}
\tilde{\redradius}_S\simeq 1+\frac{c_b}{2}\epsilon_S\log\epsilon_S\,,
\end{equation}
so that it is slightly smaller than the standard Schwartzschild radius, for a positive $c_{b}$. Of course, the potential vanishes for $\redradius\to\infty$ (and so, also for $r\to\infty$), as in the standard Schwarzschild scenario.
\begin{figure}
    \centering
    \includegraphics[width=\linewidth]{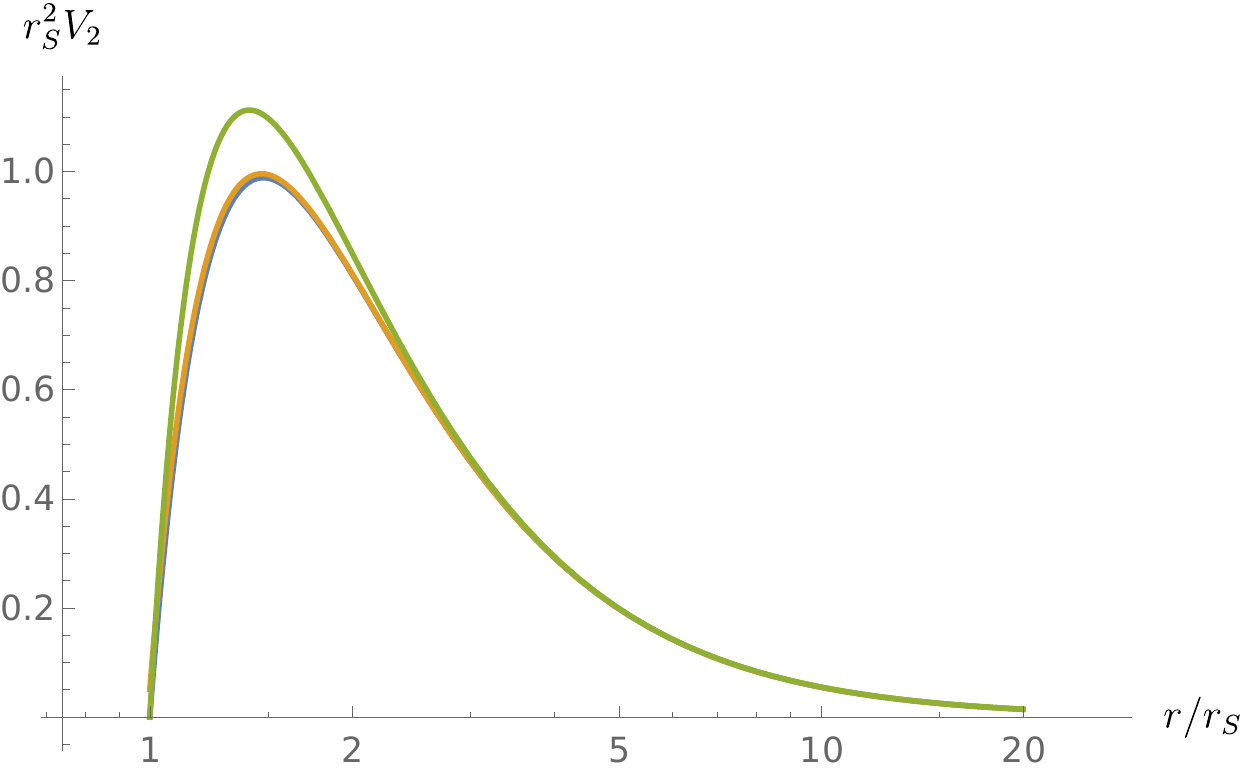}
    \caption{The potential $V_\ell(r)$ for $\ell=2$ and for different values of the $\epsilon_S$ parameter, from bottom to top: $\epsilon_S=\{10^{-38},10^{-1},1\}$, and for $b=3/2$..}
    \label{fig:plotpotential}
\end{figure}

\begin{figure}
    \centering
    \includegraphics[width=\linewidth]{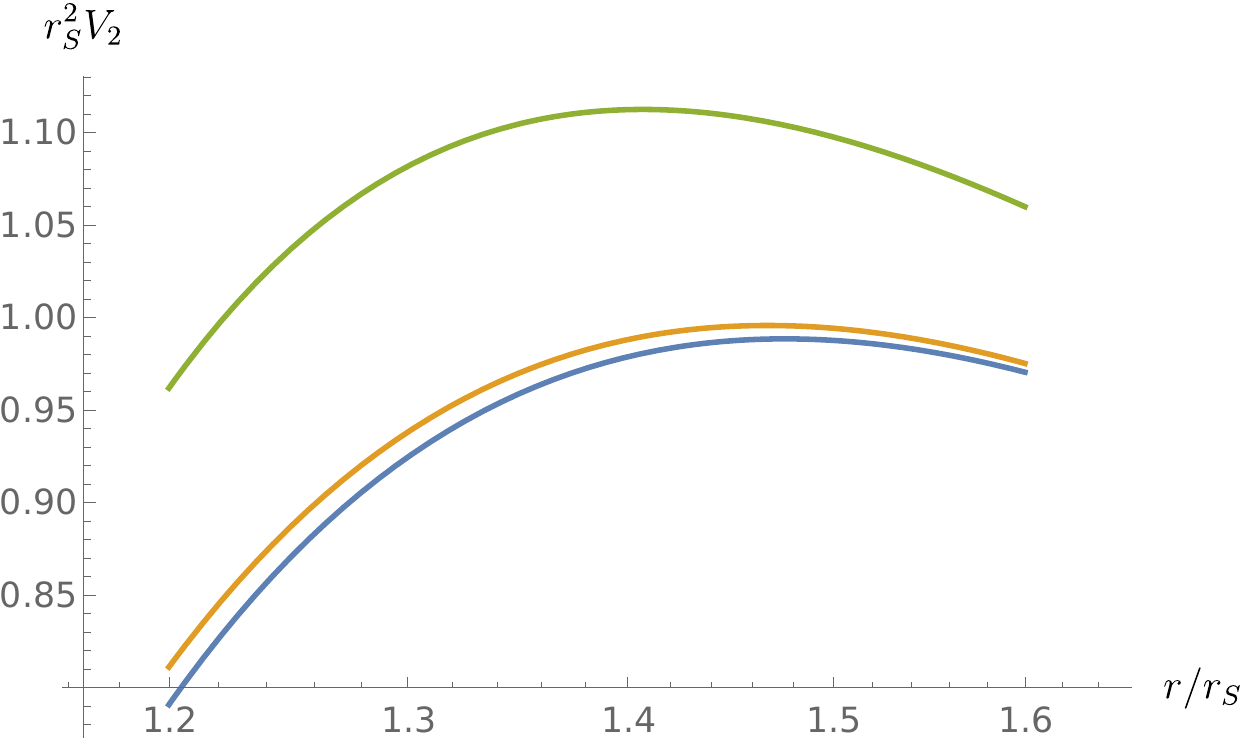}
    \caption{A zoom in of the potential $V_\ell(r)$ for $\ell=2$ around its maximum and for different values of the $\epsilon_S$ parameter, from bottom to top: $\epsilon_S=\{10^{-38},10^{-1},1\}$, and for $b=3/2$.}
    \label{fig:plotpotentialzoom}
\end{figure}

In Figure \ref{fig:plotpotential}, we have plotted the potential as a function of $\redradius$ for various values of the ratio $\epsilon_S$, from values typical of solar mass black holes and for much smaller Schwarzschild radii. We see that the form of the potential is very much the same of the $\epsilon_S=0$ case (for $\epsilon_S\ll 1$), and, as Figure  \ref{fig:plotpotentialzoom} shows, the largest differences appear around the maximum of the potential itself. The asymptotic behavior, on the other hand, is very slighly modified.

The next standard step to study perturbations on a Schwartzschild background consists in introducing the \virgolette{tortoise coordinate} via the differential relation 
\begin{equation}\label{eqn:tortoise}
    \diff x\equiv f_{b}^{-1}(r)\diff \redradius\,.
\end{equation}
In the limit of small $\epsilon_S$, one can explicitly check that $\redradius\to \tilde{\redradius}_S$ corresponds to $x\to-\infty$, while $\redradius\to+\infty$ corresponds to $x\to+\infty$. Using the (rescaled) tortoise coordinate $x$, equation \eqref{eqn:equationmotion1} becomes, after a time Fourier transform,
\begin{equation}\label{eqn:rwscalar}
    \partial_x^2 \tilde{u}_{\ell m}(\omega,x)+Q_\ell(\omega,x)\tilde{u}_{\ell m}(\omega,x)=0\,,
\end{equation}
where 
\begin{equation}\label{eqn:qpotential}
    Q_\ell(\omega,x)\equiv r_{S}^2\left(\omega^2-V_\ell(x)\right)\,.
\end{equation}

\subsubsection{WKB approximantion and QNMs}
The WKB approximation described in \cite{iyer1,iyer2}, defines the quasinormal modes via a \virgolette{Bohr-Sommerfeld quantization rule},
\begin{equation}\label{eqn:wkbquasinormal}
    \frac{Q_{\ell}(x)}{\sqrt{2\partial^2_xQ_{\ell}(x)}}\biggr\vert_{x=x_{\max}}=\textcolor{red}{-}i(n+1/2)\,,
\end{equation}
where $n$ is an integer, $n=0,1,2,\dots$ and where $x_{\max}$ denotes the value of the tortoise coordinate $x$ at which $Q_{\ell}(x)$ has a maximum. Given that the potential \eqref{eqn:vll} is modified mainly in the region around its maximum, the above approximation may provide already a good way to determine the corrections to the quasinormal modes in this modified Schwarzschild background. 

Since we will work perturbatively, keeping only terms at order $\mathcal{O}(\epsilon_S)$, we will need to expand the potential $Q_\ell(x)$ in powers of $\epsilon_S$:
\begin{equation}
    Q_\ell(x)\equiv Q_{\ell}^{(0)}(x)+\epsilon_SQ_\ell^{(1)}(x)+\mathcal{O}(\epsilon_b^2)\,.
\end{equation}
The point at which $Q_{\ell}(x)$ reaches its maximum will be slighly displaced from the point $\overline{x}_{\ell}$ at which $Q_\ell^{(0)}(x)$ is maximum, meaning that we can write $x_{\ell,\max}=\overline{x}_{\ell}+\epsilon_S x_{\ell,\epsilon}+\mathcal{O}(\epsilon_S^2)$. Then we have
\begin{align*}
     Q_\ell(x_{\ell,\max})     &=Q_{\ell}^{(0)}(\overline{x}_{\ell})\\
     &\quad+\epsilon_Sx_{\ell,\epsilon}\partial_xQ_{\ell}^{(0)}(\overline{x}_{\ell})+\epsilon_SQ_\ell^{(1)}(\overline{x}_{\ell})+\mathcal{O}(\epsilon_S^2)\\
     &=Q_{\ell}^{(0)}(\overline{x}_{\ell})+\epsilon_SQ_\ell^{(1)}(\overline{x}_{\ell})+\mathcal{O}(\epsilon_S^2)\,,
\end{align*}
since $\overline{x}_{\ell}$ is the maximum of $Q_\ell^{(0)}$. Notice, at this point, that since the map $x(\redradius)$ is monotonic for small $\epsilon_S\ll 1$, the maximum of the potential can be equivalently determined for $\redradius$, which is certainly easier:
\begin{align}\label{eqn:barql}
    \overline{Q}_\ell&\equiv Q_\ell(x_{\ell,\max})=Q_\ell(\redradius_{\ell,\max})\nonumber\\&=Q_{\ell}^{(0)}(\overline{\redradius}_\ell)+\epsilon_SQ_\ell^{(1)}(\overline{\redradius}_\ell)+\mathcal{O}(\epsilon_b^2)\nonumber\\
    &=\overline{Q}_{\ell}^{(0)}+\epsilon_b\overline{Q}_\ell^{(1)}+\mathcal{O}(\epsilon_b^2)\,,
\end{align}
 where $\redradius_{\ell,\max}\equiv \overline{\redradius}_\ell+\epsilon_b \redradius_{\ell,\epsilon}+\mathcal{O}(\epsilon_b^2)$, and $\redradius_{\ell,\epsilon}$ is determined from the requirement that $\partial_\redradius Q_\ell\vert_{\redradius=\redradius_{\max}}=0$ order by order. In this way one finds,
 \begin{equation}
     \redradius_{\ell,\epsilon}=-\frac{\partial_\redradius Q_\ell^{(1)}}{\partial^2_\redradius Q_{\ell}^{(0)}}\Biggr\vert_{\redradius=\overline{\redradius}_\ell}\,.
 \end{equation}
 Explicitly, $\overline{Q}_\ell^{(1)}$ in equation \eqref{eqn:barql} is given by
 \begin{equation}
     \overline{Q}_\ell^{(1)}=2\tilde{\omega}^{(0)}\tilde{\omega}^{(1)}-\overline{\tilde{V}}_\ell^{(1)}\,,
 \end{equation}
 where $\overline{\tilde{V}}\ell^{(1)}$ is the first order term in the expansion of $\tilde{V}_\ell(\redradius)$ in powers of $\epsilon_S$ evaluated in $\overline{\redradius}_{\ell}$, where $\tilde{V}_{\ell}(\redradius)\equiv r_S^2V_{\ell}(\redradius)$ is the potential in units of the Schwartzschild radius. Similarly, $\tilde{\omega}\equiv r_{S}\omega$ represents a frequency in units of the Schwartzschild radius.
 
The same strategy can be used for the denominator in equation \eqref{eqn:wkbquasinormal}: 
\begin{align*}
    \partial^2_x\overline{Q}_\ell&\equiv \partial^2_xQ_\ell(x)\vert_{x=x_{\ell,\max}}=\partial_\redradius^2 Q_\ell(\redradius)(\partial_x \redradius)^2\vert_{\redradius=\redradius_{\ell,\max}}\\
   &= \left[\left(\partial_\redradius^2Q_\ell^{(0)}+\epsilon_S\partial^2_\redradius Q_\ell^{(1)}+\mathcal{O}(\epsilon^2_S)\right)f_{b}^2(\redradius)\right]_{\redradius=\redradius_{\ell,\max}}
\end{align*}
The zeroth-order contribution of this equation is easily recognized:
\begin{subequations}\label{eqn:seconddq}
\begin{equation}
    \partial^2_x\overline{Q}_\ell^{(0)}=\left[\partial_\redradius^2Q_\ell^{(0)}\left(1-\redradius^{-1}\right)^2\right]_{\redradius=\overline{\redradius}_\ell}\,,
\end{equation}
while at the first order, we have
 \begin{align}\label{eqn:seconddq1}
      \partial^2_x\overline{Q}_\ell^{(1)}&=\left[\partial_\redradius^3Q_\ell^{(0)}\left(1-\redradius^{-1}\right)^2\redradius_{\epsilon}\right]_{\redradius=\overline{\redradius}_\ell}\nonumber\\
      &\quad+\left[\partial_\redradius^2Q_\ell^{(1)}\left(1-\redradius^{-1}\right)^2\right]_{\redradius=\overline{\redradius}_\ell}\nonumber\\
      &\quad+\left[\frac{2c_b}{\redradius^3}\left(1-\redradius^{-1}\right)\log\frac{\redradius}{\sqrt{\epsilon_S}}\partial_\redradius^2Q_\ell^{(0)}\right]_{\redradius=\overline{\redradius}_\ell}\,.
 \end{align}
\end{subequations}
In conclusion, the leading order WKB approximation \eqref{eqn:wkbquasinormal} provides a correction to the quasinormal modes of the form
\begin{equation}\label{eqn:perturbationsqnm}
    \tilde{\omega}_{\ell n}^{(1)}=\frac{1}{2\tilde{\omega}_{\ell n}^{(0)}}\left[\overline{\tilde{V}}_\ell^{(1)} -i(n+1/2)\frac{\partial^2_x\overline{Q}_\ell^{(1)}}{\sqrt{2\partial^2_x\overline{Q}_\ell^{(0)}}}\right],
\end{equation}
where $\tilde{\omega}_{\ell n}^{(0)}$ and $\tilde{\omega}_{\ell n}^{(1)}$ are the zeroth- and first-order contributions to the quasinormal modes $\tilde{\omega}_{\ell n}$:
\begin{equation}
    \tilde{\omega}_{\ell n}=\tilde{\omega}_{\ell,n}^{(0)}+\epsilon_S\tilde{\omega}_{\ell n}^{(1)}+\mathcal{O}(\epsilon^2_S)\,,
\end{equation}
and $\partial^2_x\overline{Q}_\ell^{(0)}$ and $\partial^2_x\overline{Q}_\ell^{(1)}$ are determined from equations \eqref{eqn:seconddq}. Equation \eqref{eqn:perturbationsqnm} gives the perturbations to the Schwartzschild black-hole as a result of the above leading order WKB approximation. Further corrections to this approximations may not be particularly relevant for any practical purpose, because of the already very large (at least for macroscopic black holes) suppression factor $\epsilon_S$.\\
\subsection{Axial tensor perturbations}\label{sec:axialpert}
Let us now investigate tensor perturbations on the above modified black hole solution. The perturbative framework consists in writing $g_{\mu\nu}=\bar{g}_{\mu\nu}+h_{\mu\nu}$, where $\bar{g}_{\mu\nu}$ is the background metric characterized by equation \eqref{eqn:fbr}, while $h_{\mu\nu}$ encodes the gravitational perturbation. Because of the background spherical symmetry, $h_{\mu\nu}$ can be decomposed in two parts: one depending only on $t$ and $r$, the other depending only on angular variables $\theta$ and $\varphi$. Moreover, since the background is also parity invariant, axial and polar components of $h_{\mu\nu}$ do not mix, and so $h_{\mu\nu}$ can be further decomposed into 
\begin{equation*}
    h_{\mu\nu}=h_{\mu\nu}^{\text{axial}}+h_{\mu\nu}^{\text{polar}}\,.
\end{equation*}
We will study these two different type of perturbations separately. In order to keep the notation as simple as possible, we will set $r_{S}\equiv 1$.

In the Regge-Wheel gauge (see \cite{Maggiore:2018sht} for a pedagogical discussion), the axial part of $h_{\mu\nu}$ depends only on two independent functions, $h^{(0)}$ and $h^{(1)}$, whose dynamics is determined by the equation\footnote{The form of this equation is not affected by small quantum corrections since it is already at first order.} $\delta R_{\mu\nu}=0$. The two resulting independent equations can be written, after a spherical harmonics decomposition and for $\ell\ge2$, as \cite{Cruz:2015bcj,Maggiore:2018sht}
\begin{subequations}
\begin{align}
   0&= -f_{b}^{-1}\partial_t h^{(0)}_{\ell m}+\partial_r\left(f_{b}h^{(1)}_{\ell m}\right),\\
   0&=-\partial_t\partial_rh^{(0)}_{\ell m}+\partial_t^2h^{(1)}_{\ell m}+\frac{2}{r}\partial_th^{(0)}_{\ell m}\label{eqn:dynamicaxial}\\
   &\quad+\left[\frac{f_{b}'}{r^2}\ell(\ell+1)-\frac{2}{r}f_{b}f_{b}'-\frac{2}{r^2}f_{b}^{2}\right]h^{(1)}_{\ell m}\nonumber\,,
\end{align}
\end{subequations}
where a prime denotes a differetiation with respect to $r$ and where $h^{(i)}=\sum_{\ell m}h^{(i)}_{\ell m}(t,r)Y_{\ell m}(\theta,\varphi)$. Notice that, by defining $\psi_{\ell m}\equiv f_{b}h^{(1)}_{\ell m}$, the first of the two equations above becomes
\begin{equation}\label{eqn:substitution}
    \partial_t h^{(0)}_{\ell m}=\partial_x \psi_{\ell m}\,,
\end{equation}
where $x$ is the tortoise coordinate defined by \eqref{eqn:tortoise}. By substituting eqation \eqref{eqn:substitution} in equation \eqref{eqn:dynamicaxial}, we immediately obtain 
\begin{align*}
    0&=-\partial_r\partial_x\psi_{\ell m}+f_{b}^{-1}\partial_t^2\psi_{\ell m}+\frac{2}{r}\partial_x\psi_{\ell m}\\
   &\quad+\left[\frac{\ell(\ell+1)}{r^2}-\frac{2}{r}f_{b}'-\frac{2}{r^2}f_{b}\right]\psi_{\ell m}\nonumber\,,
\end{align*}
The first order derivative term proprtional to $\partial_x\psi$ can be eliminated by introducing the usual rescaled variable $F_{\ell m}\equiv \psi_{\ell m}/r$. Indeed,
\begin{equation*}
    -\partial_r\partial_x\psi_{\ell m}+\frac{2}{r}\partial_x\psi_{\ell m}=-rf_{b}^{-1}\partial^2_xF_{\ell m}-F_{\ell m}f_{b}'+\frac{2}{r}F_{\ell m}f_{b}
\end{equation*}
and, therefore, we find an equation of the form
\begin{equation}\label{eqn:rwaxial}
    -\partial_x^2F_{\ell m}+\partial^2_t F_{\ell m}+U_{\ell}F_{\ell m}=0\,,
\end{equation}
where
\begin{equation}
    U_\ell=\frac{f_{b}}{r}\left[\frac{\ell(\ell+1)}{r}-3f_{b}'\right].
\end{equation}
By a Fourier transform, we see that equation \eqref{eqn:rwaxial} can be recast into an equation identical to \eqref{eqn:rwscalar}, but with a different \virgolette{potential}. In order to unify the treatment, we introduce an additional label, $\sigma$, which is equal to $0$ in the scalar case, and to $2$ in the axial tensor case \cite{Maggiore:2018sht,Berti:2009kk}. Then we can unify the dynamical equations for scalar and axial tensor perturbations in the following single equation,
\begin{equation}
    \partial_x^2F_{\ell m}(\omega,x)+Q_{\ell,\sigma}F_{\ell m}(\omega,x)=0\,,
\end{equation}
where
\begin{subequations}
\begin{align}
    Q_{\ell,\sigma}&=\omega^2-V_{\ell,\sigma}\,,\\
    V_{\ell,\sigma}&=\frac{f_{b}}{r}\left[\frac{\ell(\ell+1)}{r}+(1-\sigma^2)f_{b}'\right].
\end{align}
\end{subequations}
Of course, we can now determine again the quasinormal modes 
\begin{equation}
    \omega_{\ell n,\sigma}=\omega_{\ell n,\sigma}^{(0)}+\epsilon_b \omega_{\ell n,\sigma}^{(1)}+\mathcal{O}(\epsilon_b^2)
\end{equation}
in the leading order WKB approxmation by means of equation \eqref{eqn:perturbationsqnm} with the appropriate introduction of the label $\sigma$:
\begin{equation}
    \omega_{\ell n,\sigma}^{(1)}=\frac{1}{2\omega_{\ell n,\sigma}^{(0)}}\left[\overline{V}_{\ell,\sigma}^{(1)} -i(n+1/2)\frac{\partial^2_x\overline{Q}_{\ell,\sigma}^{(1)}}{\sqrt{2\partial^2_x\overline{Q}_{\ell,\sigma}^{(0)}}}\right],
\end{equation}
where, as usual, $\overline{V}_{\ell,\sigma}^{(1)}$ is the first order term in the expansion of $V_{\ell,\sigma}$ in powers of $\epsilon_b$ evaluated in $\overline{r}_{\ell,\sigma}$ (which is the radius at which the potential $V_{\ell,\sigma}^{(0)}$ is maximum), while 
\begin{subequations}
\begin{align}
    \partial^2_x\overline{Q}_{\ell,\sigma}^{(0)}&=\left[\partial_r^2Q_{\ell,\sigma}^{(0)}\left(1-r_S/r\right)^2\right]_{r=\overline{r}_{\ell,\sigma}}\,,\\
      \partial^2_x\overline{Q}_{\ell,\sigma}^{(1)}&=\left[\partial_r^3Q_{\ell,\sigma}^{(0)}\left(1-r_S/r\right)^2r_{\epsilon}\right]_{r=\overline{r}_{\ell,\sigma}}\\
      &\quad+\left[\partial_r^2Q_{\ell,\sigma}^{(1)}\left(1-r_S/r\right)^2\right]_{r=\overline{r}_{\ell,\sigma}}\nonumber\\
      &\quad+\left[\frac{2c_b}{r^3}\left(1-r^{-1}\right)\log\frac{r}{\sqrt{\epsilon_S}}\partial_r^2Q_\ell^{(0)}\right]_{r=\overline{r}_{\ell,\sigma}}\nonumber\,.
 \end{align}
\end{subequations}

\section{Conclusions and Discussions}
\label{sec:CONCL}

In this paper, we have derived a modified Schwarzschid-de Sitter geometry by including logaritmic corrections to the entropy-area law, motivated by many result in QG theories in presence of spherical symmetry. We have adopted the perspective that the thermodynamics is the fundamental framework to describe continuum gravitational dynamics, even when the thermodynamic functionals include quantum corrections to the classical ones. Therefore, we have assumed that the the fundamental thermodynamic relations characterizing the Schwarzschild-de Sitter spacetime remain unchanged even when quantum corrections to the area law are included and used them to obtain a modified Schwarzschild-de Sitter geometry for the region comprised between the black hole horizon radius and the de Sitter one. The resulting geometry is obtained within a perturbative approach, by only retaining the lowest corrections in powers of the ratio between the Planck scale and the black hole radius. Since only regions outside the black hole horizon are considered, no attempt to investigate (and possibly remove) the singularity has been performed. 

As a result of the analysis, the non-trivial metric components are equal to the classical ones, modulo the introduction of effective, $r$ dependent black hole and cosmological constant energy density parameters. Corrections of classical parameters are suppressed by (at least) a factor $r^{-2}$. 

Within this modified geometry one can in principle study a plethora of phenomena of interest for black hole physics. As an example, especially relevant in the context of gravitational waves interferometry, we have analythically computed the corrections to the black-hole quasi-normal modes for scalar and axial tensor perturbations on this modified Schwarzschild geometry (thus imposing the cosmological constant energy density to zero) within the WKB approximation. 

As the Regge-Wheeler potential gets similar corrections to the metric, one obtains that the corresponding corrections to the quasi-normal modes are suppressed by $\epsilon_S\equiv L_{\text{pl}}^2/r_S^2$. As a consequence, they are far from being observable with the current (and within the recent future) sensitivity of gravitational waves detectors. It is important to stress that these computations can only be seen as indications, given that black holes of astrophysical interest have in general non-zero spin and thus cannot be described in terms of a Schwarzschild geometry.

Still, the generalization of the techniques used within this work to less symmetric spacetimes can be quite complicated. The main reason is that there is still no conclusive argument from QG suggesting the specific type of entropy corrections for non-spherical surfaces. Of particular importance would be, in this context, to obtain clear indications about the possible entropy corrections to local Rindler horizons, which played a major role in determining the connection between gravitational dynamics and spacetime thermodynamics and thus could offer the possibility of obtaining modified gravitational dynamics from modified thermodynamic functionals, in the same spirit employed here and e.g.\ in \cite{Alonso-Serrano:2020dcz}. In particular, given the classical connection between horizon entropy and diffeomorphism invariance \cite{Wald:1993nt,Jacobson:2015uqa}, a better understanding of quantum corrections to horizon entropy may be useful to obtain some more information about the emergence of diffeomorphism invariance from QG in some continuum limit of the theory.  

The obvious technical and conceptual limitations of this approach are however balanced by the relative simplicity in which modified geometries can be obtained from simple theoretical evidences obtained from QG. In this sense, this work clearly testify the importance that spacetime thermodynamics may have in the quest to obtain an effective gravitational description in terms of continuos geometries.  
\section*{Acknowledgements}
The authors thank Daniele Oriti for discussions on an early draft of the paper.
LM thanks
the University of Pisa and the INFN (section of Pisa) for financial support, and the Ludwig Maximilians-Universit\"at (LMU) Munich for the hospitality.

\nocite{*}

\appendix

\section{Alternative derivation of modified solution}
\label{app:calcolointegrato}

We can rewrite Eq.~\eqref{eqn:conservationwithg} by using the modified
form of the entropy~\eqref{eq:modifiedS}
obtaining the differential equation (setting $C=0$)
\begin{align}
\frac{df_{\rho,b}}{du} & = \frac{4}{3} \frac{3 \Upsilon  L_{\text{pl}} u^{-2}\sqrt{\pi} 
	-4 L_{\text{pl}}^4 \rho u   }{ 1 + b u^{-2}\log u^2} \nonumber \\
	& \simeq \frac{4}{3} \left(  \frac{3 \Upsilon  L_{\text{pl}} \sqrt{\pi}}{u^2}
	-4 L_{\text{pl}}^4 \rho u \right) \left( 1 - \frac{b}{u^2} \log u^2 \right)
\end{align}
where we introduced the adimensional variable $u=\sqrt{\pi}r/L_{\text{pl}}$ and retained the corrections only at the first order.

Integrating we get
\begin{align}
    f_{\rho,b} &= c_1 -\frac{4 \Upsilon  L_{\text{pl}} \sqrt{\pi}}{u} \left[1- \frac{2b}{9 u^2} \left( 1+3\log u \right)   \right]
    \nonumber \\
    & - \frac{8}{3} L_{\text{pl}}^4 \rho u^2 \left[1  - \frac{2b}{u^2} \log^2 u \right]
\end{align}
We fix the arbitrary constant by imposing that the classical solution be obtained when $b=0$. In this way we get  (neglecting $1$ with respect to $\log u$)
\begin{align}
    f_{\rho,b} &= 1 -\frac{r_s}{r} \left(1- \frac{2 b L_{\text{pl}}^2}{3\pi r^2} \log \frac{\sqrt{\pi}r}{L_{\text{pl}}} \right) \nonumber \\
     &- \frac{8}{3} L_{\text{pl}}^2 \rho \pi  r^2 \left(1  - \frac{2 b L_{\text{pl}}^2}{\pi r^2} \log^2 \frac{\sqrt{\pi}r}{L_{\text{pl}}} \right)
\end{align}
which is the same result obtained in equation \eqref{eqn:fbr} in the main text.

\bibliography{modifiedbh}
\end{document}